\documentclass[preprint,aps,nofootinbib,preprintnumbers,amsmath,amssymb,11pt]{revtex4}
\usepackage{epsfig}
\graphicspath{ {images/} }
\usepackage{amsmath}
\usepackage{slashed}
\usepackage{amssymb}
\usepackage{hyperref}  
\usepackage{amsfonts}
\usepackage{graphicx}
\usepackage{dcolumn}
\usepackage{bm}
\usepackage{natbib}

\newcommand{\be}{\begin{equation}}
\newcommand{\ee}{\end{equation}}
\newcommand{\bea}{\begin{eqnarray}}
\newcommand{\eea}{\end{eqnarray}}

\newcommand{\nn}{\nonumber}

\def\cO {{\cal O}} 
\def\cN {{\cal N}} 
\def\cB {{\cal B}} 
\def\cF {{\cal F}} 
\def\s{\sigma}
\def\e{\epsilon} 
\def\l{\lambda} 
\def\De{\Delta} 
\def\im{{{\rm i}}}
\begin{document}


\title{Bootstrapping minimal $\mathcal{N}=1$ superconformal field theory in three dimensions}
\author{\vspace*{1 cm}\large
Junchen Rong$^{\,\clubsuit}$, Ning Su$^\dagger$}
\affiliation{\vspace*{0.5 cm}$^\clubsuit$ Center for Theoretical Physics of the Universe, Institute for Basic Science, 
Daejeon 34126, Korea\\}
\affiliation{$^\dagger$ CAS Key Laboratory of Theoretical Physics, Institute of Theoretical Physics, \\Chinese Academy of Sciences, Beijing 100190, China\\\\\\}

\begin{abstract}
\vspace*{1 cm}
Using numerical bootstrap method, we determine the critical exponents of the minimal three-dimensional $\mathcal{N}=1$ superconformal field theory (SCFT) to be $\eta_{\s}=0.168888(60)$ and $\omega=0.882(9)$. The model was argued in \cite{Grover:2013rc} to describe a quantum critical point (QCP) at the boundary a $3+1$D topological superconductor. More interestingly, the QCP can be reached by tuning a single parameter, where supersymmetry (SUSY) is realised as an emergent symmetry. By imposing emergent SUSY in numerical bootstrap, we find that the conformal scaling dimension of the real scalar operator $\sigma$ is highly restricted. If we further assume the SCFT to have only two time-reversal parity odd relevant operators, $\s$ and $\s'$, we find that allowed region for $\Delta_{\s}$ and $\Delta_{\s'}$ becomes an isolated island. 
The result is obtained by considering not only the four point correlator $\langle \s\s \s\s \rangle$, but also $\langle \s \e \s \e \rangle$ and $\langle \e\e\e\e \rangle$, with $\e\sim \s^2$ being the superconformal descendant of $\s$.
\end{abstract}

\vspace*{3 cm}
\maketitle

\def\thesection{\arabic{section}}
\def\thesubsection{\arabic{section}.\arabic{subsection}}
\numberwithin{equation}{section}
\newpage

\section{Introduction}
\label{Introduction}

The model we will be interested in is a three dimensional quantum field theory with the following Lagrangian,
\be\label{Lagan}
\mathcal{L}=\frac{1}{2}(\partial_{\mu} \sigma)^2+\bar{\psi} \slashed{\partial}\psi+\frac{\lambda_1}{2}\sigma\bar{\psi}{\psi}+\frac{\lambda_2^2}{8}\sigma^4.
\ee
Here $\psi$ is a Majorana spinor in three dimensions. The theory is invariant under time reversal symmetry (T-parity)  under which $\sigma\rightarrow -\sigma$ and $\psi \rightarrow \gamma^0\psi$. When $\lambda_1=\lambda_2$, the model has $\mathcal{N}=1$ supersymmetry (SUSY) and the Lagrangian can be rewritten into a Wess-Zumino model with superpotential $W=\Sigma^3$. Here $\Sigma=\sigma+\bar{\theta}\psi+\frac{1}{2}\bar{\theta}\theta \e$ is a real superfield. When $\lambda_1 \neq \lambda_2$, SUSY is broken. However, it is expected that the theory would still flow to the supersymmetric fixed point, and SUSY is realised as an emergent symmetry. It was argue in \cite{Grover:2013rc} that this fixed point might be realised as a quantum critical point at the boundary of a $3+1$D topological superconductor. The phenomenon of emergent supersymmetry has recently received a lot of attention. Especially, both the $1+1$D version of \eqref{Lagan} (the tricritical Ising model) \cite{Li:2016drh}  and the $2+1$D $\mathcal{N}=2$ super-Ising model \cite{Li:2017dkj} were realised using quantum Monte Carlo simulation, both of which show the property of emergent supersymmetry. 

In previous attempts to determine the critical exponents of \eqref{Lagan}, either based on bootstrapping three dimensional scalar operators \cite{Bashkirov:2013vya}, or fermions \cite{Iliesiu:2015qra}, the only SUSY constrain taken into account was the relation between the scaling dimension of operators, $\Delta_{\epsilon}=\Delta_{\s}+1$ and $\Delta_{\psi}=\Delta_{\s}+1/2$, since they belong to the same supermultiplet $\Sigma$. Here we show that we can further impose in numerical bootstrap the SUSY constrains for  OPE coefficients $\lambda_{\s\s  O}$, $\lambda_{\e\e  O}$ and $\lambda_{\s\e  O'}$ (when $ O$ and $ O'$ are in the same supermultiplet). Emergent supersymmetry means that the critical point can be reached by tuning a single parameter, which is crucial for experimental realisation. This condition is equivalent to requiring the spectrum to contain only one relevant scalar operator that is even under time-reversal parity (T-parity). Imposing such a condition, we can determine $\Delta_{\sigma}$ to high precision, providing both a upper and lower bound for its value. Furthermore, if we assume that there are only two relevant operators that are T-parity odd in the spectrum, the allow region for $(\De_{\s},\De_{\s'})$ becomes an isolated island. This helps us determine the critical exponents $\eta_{\s}, \eta_{\psi}, 1/\nu$ and $\omega$ all together. We also calculate the value of two point function for stress-energy tensor $C_T$.


\section{Bootstrapping $\mathcal{N}=1$ SCFT}
Four point functions of superfields $\langle \Sigma(x_1, \theta_1)\Sigma(x_2,\theta_2)\Sigma(x_3,\theta_3)\Sigma(x_4, \theta_4)\rangle$, when expanded in $\theta$, contains the four point function involving not only the superconformal $\s$, but also super-descendants $\psi$ and $\e$. For simplicity, in the work, we consider only bosonic fields $\s$ and $\epsilon$. In \cite{Kos:2014bka}, the critical exponents of the three dimensional Ising model is determined to high precision by studying the set of crossing equation involving the set of four point functions $\{\langle \sigma\sigma\sigma\sigma\rangle,\langle \sigma\sigma\epsilon\epsilon\rangle,\langle \epsilon\epsilon\epsilon\epsilon\rangle\}$. The crossing equations turn out to be 
\be\label{nonsusycrossing}
 \sum_{ O^+} \begin{pmatrix}\l_{\s\s O} & \l_{\e\e O}\end{pmatrix} \vec{V}_{+,\De,\ell}\begin{pmatrix} \l_{\s\s O} \\ \l_{\e\e O} \end{pmatrix}+ \sum_{ O^-} \l_{\s\e O}^2 \vec{V}_{-,\De,\ell}  =  0 ,
\ee 
The vectors $\vec{V}_{\pm}$ were calculated in \cite{Kos:2014bka}. For the reader's convenience, we will note their explicit expressions in Appendix \ref{SUSYblock}. When SUSY is taken into account, suppose $ O^+$ and $ O^-$ are from the same superconformal multiplet, their OPE coefficients $\l_{\s\s O^+}$,   $\l_{\e\e O^-}$ and $\l_{\s\e O^-}$ are proportional to each other. Their ratios can be determined by considering the $\theta$-expansion of  three point functions of superfields $\langle \Sigma\Sigma O\rangle$, based on the work of \cite{Park:1999cw}. Plug these OPE relations into \eqref{nonsusycrossing}, the crossing equations becomes 
 \be\label{susycrossing}
\sum_{\substack{\text{$l \in$ even}}}\lambda^2_{\cB_+} \vec{V}^{\cB_+}_{\De,l}+\sum_{\substack{\text{$l \in$ even}}}\lambda^2_{\cB_-} \vec{V}^{\cB_-}_{\De,l}+\sum_{\substack{ \text{$j-1/2 \in$ even}}}\lambda^2_{\cF_+} \vec{V}^{\cF_+}_{\De,j}+\sum_{\substack{ \text{$j-1/2\in$ odd}}}\lambda^2_{\cF_-} \vec{V}^{\cF_-}_{\De,j}=0,
\ee
$\De$ and $l$ (or $j$) are the scaling dimensions and spins of super-conformal primaries. As will be explained in Appendix \ref{SUSYblock}, there are four different types of superconformal blocks $\vec{V}^{\cB_+}, \vec{V}^{\cB_-}, \vec{V}^{\cF_+}$ and $\vec{V}^{\cF_-}$. Each corresponds to a type of super-multiplets appearing in $\Sigma\times \Sigma$ OPE. Schematically 
\be
\Sigma \times \Sigma \sim \mathcal{B}_+^{\mu_1\ldots \mu_{l}} + \bar{Q}Q \mathcal{B}_-^{\mu_1 \ldots \mu_{l}} + Q^{\alpha} \mathcal{F}_{+,\alpha}^{\mu_1 \ldots \mu_{j-1/2}}+ Q^{\alpha} \mathcal{F}_{-,\beta}^{(\mu_1 \ldots \mu_{}}\sigma^{\mu_{l+1})}_{\alpha}{}^{\beta}.
\ee
The supermultiplets contain the following component operators 
\bea
\mathcal{B}^{(l)}_{+/-} &:& [l]^{+/-}_{\De} \xrightarrow{Q}  [l\pm 1/2]_{\De+1/2} \xrightarrow{Q}  [l]^{-/+}_{\De+1},\qquad\quad \text {with }l=\text{integer},  \nn\\
 \mathcal{F}^{(j)}_{+/-} &:& [j]_{\De} \xrightarrow{Q}\begin{array}{c}
 [j-1/2]^{+/-}_{\De+1/2} \\
{[j+1/2]^{-/+}_{\De+1/2}}\\
\end{array}\xrightarrow{Q}  [j+1]_{\De+1}. \quad \text { with }j=\text{half integer}.
\eea
The plus or minus signs on the bosonic component operators denote their parity under time-reversal. The OPE coefficients in \eqref{susycrossing} are $\lambda_{\sigma\sigma  O}$, with $ O$ being the T-parity even operator in the multiplet. 

\begin{figure}[h]
\includegraphics[width=7cm]{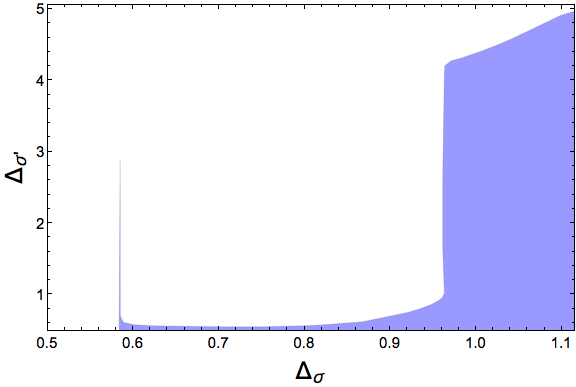}
\caption{Bound on the scaling dimension of the subleading T-parity odd operator. We assume that the spectrum contains only one relevant T-parity even operator. The numerics is performed at $\Lambda=13$.}\label{pike}
\end{figure}

Requiring the spectrum to contain only one relevant T-parity even scalar amounts to imposing the following conditions
\begin{itemize}
\item all $\mathcal{B}^{(l)}_{+}$ multiplets with $l=0$ have scaling dimension bigger than 3,
\item all $\mathcal{B}^{(l)}_{-}$ multiplets with $l=0$ (except for $\Sigma$) have scaling dimension bigger than 2,
\item all $\mathcal{F}^{(j)}_{-}$ multiplets with $j=1/2$  have scaling dimension bigger than $5/2$.
\end{itemize}
Imposing the first two conditions, assuming the sub-leading $\mathcal{B}^{(0)}_{-}$ super multiplet (which we denote the superfield as $\Sigma'$ and it super-primary as $\sigma'$) to have scaling dimension bigger or equal to $\Delta_{\sigma'}$. We can use numerical bootstrap to carve out the region in $(\Delta_{\s},\Delta_{\s'})$-plane that allows an unitary $\mathcal{N}=1$ SCFT to exist. The result is shown in Figure \ref{pike}, where the maximal derivatives for conformal block approximation is $\Lambda=13$. For details of numerical bootstrap, we refer to \cite{Simmons-Duffin:2015qma}.  The sharp spike at $\De_{\s}\approx 0.584$ indicates a SCFT which we will identify as the 3D $\cN=1$ minimal SCFT. The kink at $\Delta_{\sigma}\approx 0.96$ appears when the bound meets the line $\Delta_{\sigma'}=\De_{\sigma}$, which is possibly related to the existence of a certain $\mathcal{N}=2$ SCFT with flavor symmetries. Notice the $\cN=2$ flavor current multiplet, when viewed as $\cN=1$ multiplets, branches into a conserved current multiplet and a real scalar multiplet with the scaling dimension $\Delta=1$. 

\begin{figure}[h]
\includegraphics[width=8cm]{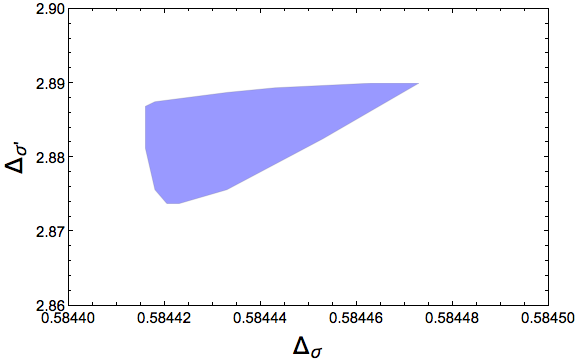}
\caption{The region that allowed a unitary $\cN=1$ SCFT to exist. We assume that the spectrum contains only one relevant T-parity even operator, and $\s$ and $\s'$ are the only two relevant T-parity odd scalar operators. The numerics is performed at $\Lambda=27$. }\label{island}
\end{figure}

If we further assume that $\Sigma$ and $\Sigma'$ are the only two relevant scalar superfields in the spectrum, the allowed region becomes an isolated island. This is shown in Figure \ref{island}, where we use the configuration $S_{\Lambda=27}$ in \cite{Simmons-Duffin:2015qma} for the numerics.
\section{low-lying spectrum}
From the island, it is easy to get $\Delta_{\s}=0.584444(30)$, corresponding to the critical exponents 
\be
\eta_{\s}=\eta_{\psi}=0.168888(60),\quad 1/\nu=1.415556(30).
\ee
$\Sigma'$ contains a super-primary with $\De_{\s'}=2.882(9)$ and also a super-descendant which is the lowest dimensional irrelevant T-parity even scalar operator. This helps us determine the critical exponent 
\be
\omega=0.882(9).
\ee 
The four loop $\epsilon$-expansion of the Gross-Neveu-Yukawa model was calculated in \cite{Zerf:2017zqi}, the Pad\'e$_{[3,1]}$ approximation gives $\eta_{\s}=\eta_{\psi}=0.170$, $1/\nu=1.415$ and $\omega=0.838$. This is consistent with our result and justifies our identification of the island with $\mathcal{N}=1$ minimal model. We can also compare our result with previous bootstrap determinations of the critical exponents. In \cite{Bashkirov:2013vya}, it was observed that the SUSY line $\Delta_{\e}=\Delta_{\s}+1$ intersects with the region at $\De_{\s}=0.565$, this provides a lower bound for $\eta_{\s}>0.13$. In \cite{Iliesiu:2015qra}, the allowed region touches the SUSY line $\De_{\psi}=\De_{\s}+1/2$ at $\Delta_{\s} \approx 0.582$ (though one need to  impose the condition $\Delta_{\sigma'}\geq 3$). Our results shows that  $\mathcal{N}=1$ minimal model is indeed located in this region.   

We can also calculate the constant $C_T$ of the stress-tensor two point function when it is normalised using the Ward identity.  By bootstrapping the OPE coefficient $\lambda^2_{\cF_-}$, with $\cF^{\De=5/2,j=3/2}_{-}$ being the SUSY current multiplet, we get
\be
C_T^{\cN=1}/C_T^{f.s.}\approx 1.684
\ee
where $C_T^{f.s.}$ means $C_T$ of a free real scalar. This value is in fair agreement with the value $C_T^{\cN=1}/C_T^{f.s.}\approx 1.73$ from one loop $\epsilon$-expansion given in \cite{Fei:2016sgs}.
 
Since we now know the critical exponents to high precision, we can use this result to perform ``two-sided'' Pad\'e resummation of the large-$N$ expansion result of Gross-Neveu model, which allows us to estimate the critical exponents of the Gross-Neveu(-Yukawa) models with higher number of fermions. The critical exponents $\eta_{\psi}$ is know to $\frac{1}{N^3}$ order \cite{Gracey:1993kc} 
\be
\eta_{\psi}=\frac{8}{3 \pi ^2 N}+\frac{1792}{27 \pi ^4 N^2}+\frac{64 \left(-3402 \zeta (3)+141 \pi ^2-668+324 \pi ^2 \log (2)\right)}{243 \pi ^6 N^3}+\mathcal{O}(\frac{1}{N^4}).
\ee
The resumed result is presented in the first two rows of Table \ref{Etapsi}, and compared with result from previous estimation using other methods. The $N=8$ model describes the quantum critical point of the semimetal to charge density wave order transition in graphene \cite{Herbut:2006cs}. Since the large-$N$ expansion of critical exponents $\nu$ and $\eta_{\s}$ is only known up to order $\frac{1}{N^2}$, we will not try to re-sum them here. 

\begin{table}[h]
\begin{tabular}{l|c|c}\hline\hline
$N$                   & 4 & 8 \\\hline
large-$N$, Pad\'e$_{[2,2]}$   & 0.0942  &  0.0430 \\ \hline
large-$N$, Pad\'e$_{[3,1]}$    & 0.1043  &  0.0437 \\ \hline
4$-\epsilon$, $\epsilon^4$,  Pad\'e$_{[2,2]}$   \cite{Zerf:2017zqi}     &  0.0976  &  0.0539 \\\hline
    2+$\epsilon$, $\epsilon^4$,  Pad\'e                  \cite{Gracey:2016mio}   &  - &  0.082\\\hline
\end{tabular}
\caption{$\eta_{\psi.}$}\label{Etapsi}
\end{table}

\section{Discussion}
A Monte Carlo simulation of this $\mathcal{N}=1$ super-Ising models would certainly be interesting. The Lagrangian \eqref{Lagan} has anomaly under time-reversal symmetry. One would possibly need to study such a system at the boundary of a $3+1$ dimensional lattice, which can make the simulation time-consuming. Another interesting direction, as pointed out in \cite{Li:2017dkj}, is to construct lattice models with non-local interaction.   It is of course also interesting to extend our work to study $\mathcal{N}=1$ SCFT with flavor symmetries. Quite recently \cite{Benini:2018umh,Gaiotto:2018yjh,Benini:2018bhk}, $\mathcal{N}=1$ QED$_3$ coupled to certain number of scalar multiplets has been shown to be dual to $\mathcal{N}=1$ Wess-Zumino model with certain flavor symmetries. It is expected that our new way of bootstrapping $\mathcal{N}=1$ SCFT would be able to reveal new features of these fixed points. 
Notice that we have neglected the four point functions involving the fermionic operator $\psi$ in our calculation, it would be interesting consider them in the future and see how they can improve our numerical bootstrap bounds.  
Another lesson that we have learnt from this work is that the four point functions involving super-descendant operators (which are conformal primaries) play important role in numerical bootstrap. It would be interesting to try imposing similar constrains when bootstrapping SCFT's with higher number of supercharges, extending the works of \cite{Bobev:2015vsa,Chester:2014fya}.

\vskip 0.4 in
{\noindent\large  \bf Acknowledgments}
\vskip 0.in

We are grateful to Ziyang Meng, Yuan Wan, Jeong-Hyuck Park, David Simmons-Duffin and Alessandro Vichi for helpful discussion or comments. J. R. would like to thank the hospitality of Institute of Physics, Chinese Academy of Science while part of the work was finished. N.S.'s research is supported in part by the Key Research Program of Frontier Sciences of CAS under Grant No. QYZDB-SSW-SYS014 and Peng Huanwu center under Grant No. 11747601. The numerical computations in this work are partially supported by HPC Cluster of SKLTP/ITPCAS. To calculate the conformal block functions, we used the code from JuliBootS program\cite{Paulos:2014vya}. The numerics is solved using SDPB program\cite{Simmons-Duffin:2015qma}.

\appendix 
\section{Superconformal Blocks}\label{SUSYblock}
We follow the result \cite{Park:1999cw} to consider the three point function involving two $\Sigma$'s and a third superfield, we get
\bea
&&\langle \mathcal{O}^{(l)}(x_1,\theta_1,\eta_1) \Sigma(x_2,\theta_2)\Sigma(x_3,\theta_3)\rangle=\frac{t({\rm X}_1,\Theta_1,\eta_1)}{x_{12}^{2\Delta_{\Phi}-\De_{\cO}-l}x_{13}^{2\Delta_{\Phi}-\De_{\cO}-l}x_{23}^{\De_{\cO}+l}},\\
&& x_{12}^{\mu}=x_1^{\mu}+x_2^{\mu}+\im \bar{\theta}_1\gamma_{\mu}\theta_2,\quad {\rm x}_{12\pm}=x_{12}^{\mu}\gamma_{\mu}\pm\im\frac{1}{2}\bar{\theta}_{12}{\theta_{12}},\quad \theta_{12}=\theta_{1}-\theta_{2},\\
&&{\rm X}_{1}=\frac{1}{2}({\rm x}_{31+}^{-1}{\rm x}_{23-}{\rm x}_{21+}^{-1}+{\rm x}_{21+}^{-1}{\rm x}_{23+}{\rm x}_{31-}^{-1}),\quad \Theta_1=\im({\rm x}_{21+}^{-1}\theta_{21}-{\rm x}_{31+}^{-1}\theta_{31}).
\eea
Here $\eta_1$ is an auxiliary two component spinor, valued by commuting numbers. There exist four different $t$-structures
\bea
\mathcal{B}^{(l)}_+&:& (\bar{\eta}_1 {\rm X}_1 \eta_1)^l,\nn\\
\mathcal{B}^{(l)}_-&:& \bar{\Theta}_1\Theta_1 (\bar{\eta}_1 {\rm X}_1 \eta_1)^l(\text{tr}[{\rm X}_{1}^2])^{-1/2},\nn\\
\mathcal{F}^{(l)}_+&:& \bar{\eta}_1{\rm X}_1\Theta_1 (\bar{\eta}_1 {\rm X}_1 \eta_1)^{l-1/2} (\text{tr}[{\rm X}_{1}^2])^{-3/4}, \nn\\
\mathcal{F}^{(l)}_-&:& \bar{\eta}_1\Theta_1 (\bar{\eta}_1 {\rm X}_1 \eta_1)^{l-1/2}  (\text{tr}[{\rm X}_{1}^2])^{-1/4} \nn.
\eea
Expanding the three point functions in series of $\theta_1$, $\theta_2$ and $\theta_3$, we obtain the the SUSY OPE relations.  It is also necessary to take into account the fact that the operators from $\theta$ expansion are not yet properly normalised. One need to fix this by expanding the two point functions 
\be
\langle \cO^{(l)}(x_1,\theta_1,\eta_1) {\cO}^{(l)}(x_2,\s_2,\eta_2)\rangle=\frac{({\bar{\eta}_1\rm x_{12+}\eta_2})^{2l}}{(x^2_{12})^{\Delta+l}},
\ee  
then reading out the operators normalization and scaling the OPE relations properly. The convention in this work is same as \cite{Poland:2018epd}. The calculation is in the same spirit of \cite{Poland:2010wg}, where $\mathcal{N}=1$ superconformal block in four dimensions was calculated.

The superfield $\cB^{(l)}_+$ can be expanded in $\theta$ as $\cB^{(l)}_+=O_+^{(l)}+\ldots+\bar{\theta}\theta (O_{-}+ P^2 O_+^{(l)})$, where the dots denote fermionic operators that we will neglect. $P^2 O_+$ is the conformal descendant of $O_+^{(l)}$. From the $\theta$-expansion of $\langle \cB^{(l)}_+ (x_1,\theta_1) \Sigma(x_2,\theta_2)\Sigma(x_3,\theta_3) \rangle$, we pick the terms promotional to $1$, $\bar{\theta}_2\theta_2\bar{\theta}_3\theta_3$ and  
 $\bar{\theta}_1\theta_1\bar{\theta}_3\theta_3$, delete the contribution of $P^2 O_+^{(l)}$, and read out the OPE ratios $\lambda_{\s\e O_-}/\lambda_{\s\s O_+}$ and $\lambda_{\e\e O_+}/\lambda_{\s\s O_+}$. To take care of the normalization, we consider two point function $\langle \cB^{(l)}_+(x_1,\theta_1)\cB^{(l)}_+(x_2,\theta_2)\rangle$. The term proportional to $1$ gives us $\langle O_+ O_+\rangle$, while the $\bar{\theta}_1\theta_1\bar{\theta}_2\theta_2$ term, after deleting $\langle P^2O_+  P^2O_+\rangle$ contribution, gives us $\langle O_- O_-\rangle$.

The $\theta$-expansion of $\langle \cB^{(l)}_{-} \Sigma\Sigma \rangle$ with $\cB^{(l)}_-=O_-^{(l)}+\ldots+\bar{\theta}\theta (O_{+}+ P^2 O_-^{(l)})$, after normalization, gives us the OPE ratios $\lambda_{\s\e O_-}/\lambda_{\s\s O_+}$ and $\lambda_{\e\e O_+}/\lambda_{\s\s O_+}$. This time, we need to study the terms proportional to $\bar{\theta}_1\theta_1$, $\bar{\theta_3}\theta_3$ and  $\bar{\theta_1}\theta_1\bar{\theta_2}\theta_2\bar{\theta_3}\theta_3$.

The fermionic superfield $\cF^{(j)}_+$ can be expanded in $\theta$ as $\cF^{(j)}_+=\ldots+ \bar{\eta}\theta \cdot O^{(l)}_{+}+ \bar{\eta}\gamma_{\mu}\theta\cdot O^{(l+1)}_{-}+\ldots$, where $l=j-1/2$. The $\bar{\eta}_1\theta_1$, $\bar{\eta}_1\theta_1\bar{\theta_2}\theta_2\bar{\theta}_3\theta_3$ and $\bar{\eta}_1\gamma_{\mu}\theta_1\bar{\theta_3}\theta_3$ terms of $\langle \cF^{(l)}_{+} (x_1,\theta_1) \Sigma(x_2,\theta_2)\Sigma(x_3,\theta_3) \rangle$ help us obtain $\lambda_{\s\e O^{(l+1)}_-}/\lambda_{\s\s O^{(l)}_+}$ and $\lambda_{\e\e O^{(l)}_+}/\lambda_{\s\s O^{(l)}_+}$. To normalise the operators, one need to get the $\bar{\eta}_1\theta_1 \bar{\eta}_2\theta_2$ and $\bar{\eta}_1\gamma_{\mu}\theta_1 \bar{\eta}_2\gamma_{\nu}\theta_2$ terms of $\langle \cF_{+}(x_1,\theta_1,\eta_1)\cF_{+}(x_2,\theta_2,\eta_2)\rangle$.

A similar calculation can be done for $\cF^{(j)}_-=\ldots+ \bar{\eta}\theta \cdot O^{(l)}_{-}+ \bar{\eta}\gamma_{\mu}\theta\cdot O^{(l+1)}_{+}$+\ldots. This time we need the  $\bar{\eta}_1\theta_1\bar{\theta}_{3}\theta_3$, $\bar{\eta}_1\gamma_{\mu}\theta_1\bar{\theta_2}\theta_2\bar{\theta_3}\theta_3$ and $\bar{\eta}_1\gamma_{\mu}\theta_1\bar{\theta_2}$ terms of $\langle \cF^{(j)}_- \Sigma \Sigma \rangle$.

Notice an interesting feature in the calculation is that only one conformal primary of the super multiplet appears in the $\s\times \s$ OPE (or $\s \times \e$ OPE). 

Plug the OPE relations into the non-SUSY crossing equation \eqref{nonsusycrossing}, where from \cite{Kos:2014bka}, we have 
\bea
\vec{V}_{-,\De,\ell} = \begin{pmatrix} 0  \\ 0 \\ F_{-,\De,\ell}^{\s\e,\s\e}(u,v) \\ (-1)^{\ell} F_{-,\De,\ell}^{\e\s,\s\e}(u,v) \\ - (-1)^{\ell} F_{+,\De,\ell}^{\e\s,\s\e}(u,v) \end{pmatrix}, && 
\vec{V}_{+,\De,\ell} = \begin{pmatrix} \begin{pmatrix}  F^{\s\s,\s\s}_{-,\De,\ell}(u,v) & 0 \\ 0 & 0  \end{pmatrix} \\ \begin{pmatrix}  0 & 0 \\ 0 & F^{\e\e,\e\e}_{-,\De,\ell}(u,v)  \end{pmatrix}\\ \begin{pmatrix}  0 & 0 \\ 0 & 0  \end{pmatrix}  \\ \begin{pmatrix}  0 & \frac12 F^{\s\s,\e\e}_{-,\De,\ell}(u,v) \\ \frac12 F^{\s\s,\e\e}_{-,\De,\ell}(u,v) & 0 \end{pmatrix} \\ \begin{pmatrix} 0 & \frac12 F^{\s\s,\e\e}_{+,\De,\ell}(u,v) \\ \frac12 F^{\s\s,\e\e}_{+,\De,\ell}(u,v) & 0  \end{pmatrix} \end{pmatrix}.\nn\\
\eea

We get the SUSY crossing equation \eqref{susycrossing}, with
\bea
&&\vec{V}^{\cB_+}_{\De,l}=\left(
\begin{array}{c}
 F_{-,\Delta,l}{}^{\s\s ,\s\s } \\
 c_1^2 F_{-,\Delta,l}{}^{{\e\e},{\e\e}} \\
 c_2F_-{}^{{\s\e},{\s\e}}_{,\Delta+1,l} \\
 c_1 F_-{}^{\s\s ,{\e\e}}_{,\Delta,l}+c_2(-1)^lF_-{}^{{\e\s },{\s\e}}_{,\Delta+1,l}\\
 c_1 F_+{}^{\s\s ,{\e\e}}_{,\Delta,l}-c_2(-1)^lF_+{}^{{\e\s },{\s\e}}_{,\Delta+1,l} \\
\end{array}
\right),\qquad\text{ }
\vec{V}^{\cB_-}_{\De,l}=\left(
\begin{array}{c}
 F_-{}^{\s\s ,\s\s }_{,\Delta+1,l} \\
 d_1^2 F_-{}^{{\e\e},{\e\e}}_{,\Delta+1,l}  \\
 d_2F_-{}^{{\s\e},{\s\e}}_{,\Delta,l} \\
 d_1F_-{}^{\s\s ,{\e\e}}_{,\Delta+1,l}  +d_2(-1)^lF_-{}^{\e\s ,{\s\e}}_{,\Delta,l} \\
 d_1F_+{}^{\s\s ,{\e\e}}_{,\Delta+1,l}  -d_2(-1)^lF_+{}^{{\e\s },{\s\e}}_{,\Delta,l} \\
\end{array}
\right)\nonumber\\
&&\vec{V}^{\cF_+}_{\De,j}=\left(\begin{array}{c}
 F_-{}^{\s\s ,\s\s }_{,\De_l,l} \\
 f_1^2F_-{}^{{\e\e},{\e\e}}_{,\De_l,l}  \\
 f_2F_-{}^{{\s\e},{\s\e}}_{,\Delta_{l},l+1} \\
 f_1F_-{}^{\s\s ,{\e\e}}_{,\De_l,l}  +f_2(-1)^{l+1} F_-{}^{\e\s ,{\s\e}}_{\De_l,l+1} \\
 f_1F_+{}^{\s\s ,{\e\e}}_{,\De_l,l}  -f_2(-1)^{l+1} F_+{}^{{\e\s },{\s\e}}_{,\De_l,l+1} \\
\end{array}\right), \quad \vec{V}^{\cF_-}_{\De,j}=\left(\begin{array}{c}
 F_-{}^{\s\s ,\s\s }_{,\De_l,l+1} \\
 e_1^2F_-{}^{{\e\e},{\e\e}}_{,\De_l,l+1}  \\
 e_2F_-{}^{{\s\e},{\s\e}}_{,\De_l,l} \\
 e_1F_-{}^{\s\s ,{\e\e}}_{,\De_l,l+1}  +e_2(-1)^l F_-{}^{\e\s ,{\s\e}}_{,\De_l,l} \\
 e_1F_+{}^{\s\s ,{\e\e}}_{,\De_l,l+1}  -e_2(-1)^l F_+{}^{{\e\s },{\s\e}}_{,\De_l,l} \\
\end{array}\right)\nn\\
\eea
For the fermionic superblocks, we define $\De_l=\De+1/2$, $l=j-1/2$. The constants
with 
\bea
&&  c_1=\frac{\left(2 \Delta_\s-\Delta -l-1\right) \left(2 \Delta_\s-\Delta +l\right)}{2 \Delta_\s \left(2 \Delta_\s-1\right)}, \quad\quad\quad\quad \text{ }   c_2=\frac{(\Delta -1) (\Delta -l-1) (\Delta +l)}{4 (2 \Delta -1) \Delta_\s \left(2 \Delta_\s-1\right)},\nn\\
&&  d_1=\frac{\left(2 \Delta_\s+\Delta -l-3\right) \left(2 \Delta_\s+\Delta +l-2\right)}{2 \Delta_\s \left(2 \Delta_\s-1\right)},\quad\quad \text{ } \text{ } d_2=\frac{(2 \Delta -1) (\Delta -l-1) (\Delta +l)}{(\Delta -1) \Delta_\s \left(2 \Delta_\s-1\right)},\nn\\
&&f_1=\frac{\left(-2 \Delta_\s-\Delta +l+4\right) \left(-2 \Delta_\s+\Delta +l+1\right)}{2 \Delta_\s \left(2 \Delta_\s-1\right)},\quad f_2=\frac{(2 l+1) (\Delta -l-2) (\Delta +l)}{2 (l+1) \Delta_\s \left(2 \Delta_\s-1\right)},\nn\\
&&e_1=\frac{\left(2 \Delta_\s-\Delta +l+1\right) \left(2 \Delta_\s+\Delta +l-2\right)}{2 \Delta_\s \left(2 \Delta_\s-1\right)}, \quad\quad\text{ } \text{ } e_2=\frac{(l+1) (\Delta -l-2) (\Delta +l)}{2 (2 l+1) \Delta_\s \left(2 \Delta_\s-1\right)}.\nn
\eea
It turns out the five crossing equations are not linear independent, when doing numerical bootstrap, we simply delete the second line.  

\end{document}